\documentclass{article}

\usepackage{float}

\usepackage{arxiv}

\usepackage[utf8]{inputenc} 
\usepackage[T1]{fontenc}    
\usepackage{hyperref}       
\usepackage{url}            
\usepackage{booktabs}       
\usepackage{amsfonts}       
\usepackage{nicefrac}       
\usepackage{microtype}      

\usepackage{authblk}

\usepackage{array}
\usepackage{graphicx}
\usepackage{esvect}
\usepackage{amsmath}

\DeclareMathOperator*{\argminA}{arg\,min}
\DeclareMathOperator*{\argmaxA}{arg\,max}

\usepackage{setspace}
\title{Benchmarking the Performance of Bayesian Optimization across Multiple Experimental Materials Science Domains}

\author[1]{\textbf{Qiaohao Liang}}
\author[2]{\textbf{Aldair E. Gongora}}
\author[3]{\textbf{Zekun Ren}}
\author[4]{\textbf{Armi Tiihonen}}
\author[4]{\textbf{Zhe Liu}}
\author[4]{\textbf{Shijing Sun}}
\author[5]{\textbf{James R. Deneault}}
\author[6]{\textbf{Daniil Bash}}
\author[7]{\textbf{Flore Mekki-Berrada}}
\author[7]{\textbf{Saif A. Khan}}
\author[6]{\textbf{Kedar Hippalgaonkar}}
\author[5]{\textbf{Benji Maruyama}}
\author[2]{\textbf{Keith A. Brown}}
\author[8]{\textbf{John Fisher III}}
\author[4]{\textbf{Tonio Buonassisi}}
\affil[1]{Department of Materials Science, Massachusetts Institute of Technology}
\affil[2]{Department of Mechanical Engineering, Boston University}
\affil[3]{Singapore-MIT Alliance for Research and Technology}
\affil[4]{Department of Mechanical Engineering, Massachusetts Institute of Technology}
\affil[5]{Air Force Research Laboratory}
\affil[6]{Agency for Science, Technology and Research (A*STAR)}
\affil[7]{National University of Singapore}
\affil[8]{Computer Science \& Artificial Intelligence Lab, Massachusetts Institute of Technology}

\begin{document}

\maketitle
\vspace{-10pt}
\begin{abstract}
\vspace{-5pt}

In the field of machine learning (ML) for materials optimization, active learning algorithms, such as Bayesian Optimization (BO), have been leveraged for guiding autonomous and high-throughput experimentation systems. However, very few studies have evaluated the efficiency of BO as a general optimization algorithm across a broad range of experimental materials science domains. In this work, we evaluate the performance of BO algorithms with a collection of surrogate model and acquisition function pairs across five diverse experimental materials systems, namely carbon nanotube polymer blends, silver nanoparticles, lead-halide perovskites, as well as additively manufactured polymer structures and shapes. By defining acceleration and enhancement metrics for general materials optimization objectives, we find that for surrogate model selection, Gaussian Process (GP) with anisotropic kernels (automatic relevance detection, ARD) and Random Forests (RF) have comparable performance and both outperform the commonly used GP without ARD. We discuss the implicit distributional assumptions of RF and GP, and the benefits of using GP with anisotropic kernels in detail. We provide practical insights for experimentalists on surrogate model selection of BO during materials optimization campaigns.






\end{abstract}

\section*{Introduction}
\vspace{-5pt}
Autonomous experimental systems have recently emerged as the new frontier for accelerated materials research. These systems excel at optimizing materials objectives, e.g. environmental stability of solar cells or toughness of 3D printed mechanical structures, that are typically costly, slow, or difficult to simulate and experimentally evaluate. While autonomous experimental systems are often associated with high sample synthesis rates via high-throughput experiments (HTE), they may also utilize closed-loop feedback from machine learning (ML) during materials property optimization. The latter has motivated integration of advanced lab automation components with ML algorithms. Specifically, active learning \cite{settles2009active, cohn1996active} algorithms have traditionally been applied to minimizing total experiment costs while maximizing machine learning model accuracy through hyperparameter tuning. Their primary utility for materials science research, where experiments remain relatively costly, lies in an iterative formulation that proposes targeted experiments with regard to a specific design objective based on prior experimental observations. Bayesian optimization (BO) \cite{shahriari2015taking, rasmussen2010gaussian, frazier2018tutorial}, one class of active-learning learning methods, utilizes surrogate model to approximate a mapping from experiment parameters to an objective criterion, and provides optimal experiment selection when combined with an acquisition function. BO has been shown to be a data-efficient closed-loop active learning method for navigating complex design spaces \cite{shahriari2015taking, springenberg2016bayesian, brochu2010tutorial, frazier2016bayesian, eriksson2019scalable, wang2017batched}. Consequently, it has become an appealing methodology for accelerated materials research and optimizing material properties \cite{solomou2018multi, yamawaki2018multifunctional, bassman2018active, rouet2016optimisation, xue2016accelerated, chang2020efficient, macleod2020self, eyke2021toward, hase2019next, ren2018accelerated, nikolaev2016autonomy, herbol2018efficient} beyond state-of-the-art.

The materials science community has seen successful demonstrations in performing materials optimization via autonomous experiments guided by BO and its variants \cite{macleod2020self, sun2021data,gongora2020bayesian,hase2018phoenics, gongora2021using, langner2020beyond}. Naturally, previous work emphasized the ability to achieve materials optimization with fewer experimental iterations. There have been very few quantitative analyses of the acceleration or enhancement resulting from applying BO algorithms and discussions on sensitivity of BO performance to surrogate model and acquisition function selection. Rohr et al. \cite{rohr2020benchmarking}, Graff et al. \cite{graff2021accelerating} and Gongora et al. \cite{gongora2020bayesian} have evaluated the performance of BO using multiple surrogate models and acquisition functions within specific electrocatalyst, ligand and mechanical structure design spaces respectively. However, comprehensive benchmarking of the performance of BO algorithms across a broad array of experimental materials systems, as we present here, has not been done. Although one could test BO across various analytical functions or emulated materials design spaces \cite{hase2021olympus, hase2018phoenics}, empirical performance evaluation on a broader collection of experimental materials science data is still necessary to provide practical guidelines. Optimization algorithms need systematic and comprehensive benchmarks to evaluate their performance, and lack of these could significantly slow down advanced algorithm development, eventually posing obstacles for building full autonomous platforms. Presented below, the benchmarking framework, practical performance metrics, datasets collected from realistic noisy experiments, and insights derived from side-by-side comparison of BO algorithms will allow researchers to evaluate and select their optimization algorithm before deploying it on autonomous research platforms. Our work provides comprehensive benchmarks for optimization algorithms specifically developed for autonomous and high-throughput experimental materials research. Ideally, it provides insight for designing and deploying Bayesian optimization algorithms that suit the sample generation rate of future autonomous platforms and tackle materials optimization in more complex design spaces.

In this work, we benchmark the performance of BO across five different experimental materials science datasets, optimizing properties of carbon nanotube polymer blends, silver nanoparticles, lead-halide perovskites, and additively manufactured polymer structures and shapes. We utilize a pool-based active learning framework to approximate experimental materials optimization processes. We also adapt metrics such as enhancement factor and acceleration factor to quantitatively compare performances of BO algorithms against that of a random sampling baseline. We observe that when paired with the same acquisition functions, Random Forest (RF) \cite{pedregosa2011scikit, breiman2001random, liaw2002classification} as a surrogate model can compete with Gaussian Process (GP) \cite{rasmussen2010gaussian} with automatic relevance detection (ARD) \cite{neal2012bayesian} that has anisotropic kernels. They also both outperform commonly used GP without ARD. Our discussion on the differences in the implicit distributional assumptions of surrogate models and the benefits of using GP with anisotropic kernels yield deeper insights regarding surrogate model selection for materials optimization campaigns. These discussions suggest guidelines on using BO for general materials optimization. We also offer open source implementation of benchmarking code and datasets to support future development of such algorithms in the field.

\vspace{-7.5pt}

\section*{Results}
\vspace{-7.5pt}

\subsection*{Experimental materials datasets}

As seen in Table 1, we have assembled a list of five materials datasets with varying sizes, dimensions $n_{dim}$, and materials systems. These diverse datasets are generated from autonomous experimental studies of collaborators, and facilitate BO performance analysis across a broad range of materials. They contain three to five independent input features, one property as materials optimization objective, and contain from a few tens to hundreds of data points.  Based on their optimization objectives, the design space input features in the datasets range from materials compositions to synthesis processing parameters, as seen in Table T1 - T5 in supplementary information (SI). 

\newcolumntype{P}[1]{>{\centering\arraybackslash}p{#1}}

\begin{table}[!h]
  \caption{Description of experimental materials science datasets.}
  \centering
  \begin{tabular}{c c c c P{1cm} P{4cm}}
    \toprule
    Dataset & Domain & Synthesis & Size & $n_{dim}$ & Optimization Objective\\
    \midrule
    P3HT/CNT \cite{bash2020machine} & Composite blends & Drop casting & 178 & 5 &  Electrical conductivity\\
    AgNP \cite{mekki2021two} & Silver nanoparticles & Flow synthesis & 164 & 5 &   Absorbance spectrum score\\
    Perovskite \cite{sun2021data} & Thin film perovskite & Spin coating & 94 & 3 &  Stability score\\
    Crossed barrel \cite{gongora2020bayesian} & 3D printed structure & 3D printing  & 600 & 4 &  Mechanical toughness\\
    AutoAM \cite{deneault2021toward} & Materials manufacturing & 3D printing  & 100 & 4 &  Shape score\\
    \bottomrule
  \end{tabular}
\end{table}



The datasets were processed for comparison purposes: 
\begin{enumerate}
  \item For each each dataset, its optimization objective values are independently centered to its mean and scaled to unit variance.
  \item For each each dataset, its optimization problems is formulated as global minimization for consistency.
\end{enumerate}

It should be noted that while all datasets were gathered from relatively high-throughput experimental systems, P3HT/CNT, AgNP, Perovskite, and AutoAM had BO guiding the selection of subsequent experiments partially through the materials optimization campaigns. Across the datasets, the differences in distribution of normalized objective values can be observed in Figure 1(a); the differences in distribution of sampled data points in its respective materials design space can be seen in Figure 1(b). The five materials datasets in the current study are available in the following GitHub repository \cite{benchmarkingGithub}.

\begin{figure}[!h]
  \centering
  \includegraphics[width = \columnwidth]{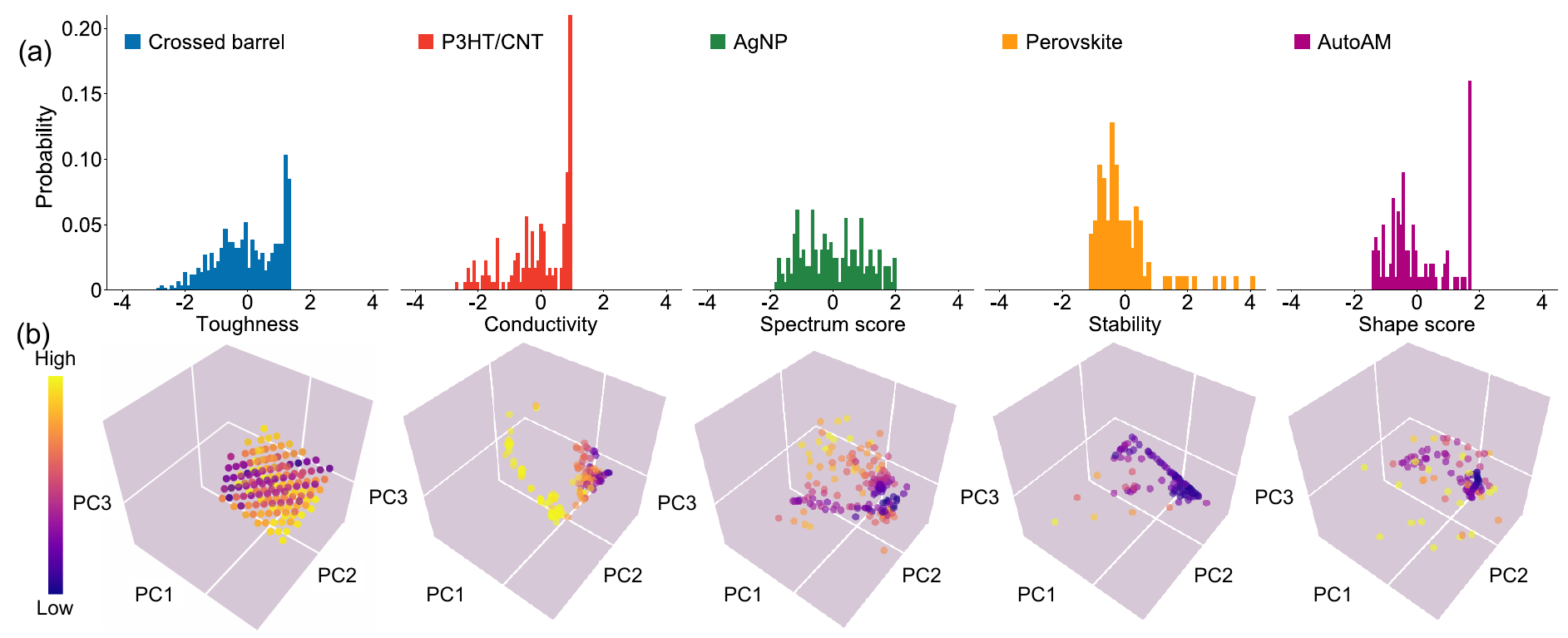}
  \caption{Experimental materials dataset design space manifold complexity visualization. (a) Histogram of objective values normalized to zero-mean without loss of generality. (b) Input feature space, i.e. design space, visualization after dimension reduction to 3D via principal component analysis (PCA). The colors of each point in the datasets indicate its value. PCA was performed to reduce all datasets dimensions to three for visualization, and the three axes shown are the top three principal component directions of each dataset.}
\end{figure}

\vspace{-8pt}

\subsection*{Bayesian Optimization: surrogate models and acquisition functions}
\vspace{-7.5pt}

Bayesian optimization (BO) \cite{shahriari2015taking, rasmussen2010gaussian, frazier2018tutorial} aims to solve the problem of finding a global optimum (min or max) of an unknown objective function $g$: $\vv{x}^*= \argminA_x g(\vv{x})$ where $\vv{x}\in X$ and $X$ is a domain of interest in $\mathcal{R}^{n_{dim}}$. BO holds the assumption that this black-box function $g$ can be evaluated at any $\vv{x}\in X$ and the responses are noisy point-wise observations ($\vv{x}$, $y$), where ${E}[y |g(\vv{x})] = g(\vv{x})$.  The surrogate model $f$ is probabilistic and consists of a prior distribution that approximates the unknown objective function $g$, and is sequentially updated with collected data to yield a Bayesian posterior belief of $g$. Decision policies aimed to find the optimum in fewer experiments are implemented in acquisition functions, which can use the mean and variance predicted at any $\vv{x}\in X$ in the posterior to select the next observation to be performed.

The BO algorithm is comprised of both a surrogate model and an acquisition function. The surrogate models considered in this study are random forest (RF) \cite{pedregosa2011scikit}, Gaussian process (GP) regression \cite{gpy2012gaussian}, and GP with automatic relevance detection (ARD) \cite{gpy2012gaussian, frazier2018tutorial, neal2012bayesian}.

\begin{enumerate}
    \item To approximate the experience of a researcher with little prior knowledge of a materials design space, for RF, we have hyperparameters applicable across all five datasets without loss of generality: $n_{\text{tree}} =$ 50 and bootstrap $=$ True.
    \item For hyperparameters of GP, we choose kernels from Mat\'ern52, Mat\'ern32, Mat\'ern12, radial basis function (RBF), and multilayer perceptron (MLP). The initial lengthscale for each kernel was set to unit length.
    \item For hyperparameters of GP ARD, we use ARD, which allows GP to keep anisotropic kernels. The kernel function of GP then has individual characteristic lengthscales $l_{j}$ \cite{frazier2018tutorial, neal2012bayesian} for each of the input feature dimensions $j$. 
\end{enumerate}

As an example, in dimension $j$, Mat\'ern52 kernel function between two points $\vv{p}, \vv{q}$ in design space would be \begin{eqnarray}
k(p_{j}, q_{j}) = \sigma_{0}^2\cdot(1 + \dfrac{\sqrt{5}r}{l_{j}} + \dfrac{5r^2}{3l_{j}^2} )\exp{(-\dfrac{\sqrt{5}r}{l_{j}})}
\end{eqnarray}
where $r = \sqrt{(p_{j} - q_{j})^2}$, $\sigma$ is standard deviation and $l_{j}$ is the characteristic lengthscale. These characteristic lengthscales can be used to estimate the distance moved along $j^{th}$ dimension from the input values in design space before the change of objective values become uncorrelated with this feature. $\dfrac{1}{l_{j}}$ is thus useful in understanding the sensitivity of objective value to input feature $j$.

We then pair the selected surrogate model with one of three acquisition functions, including expected improvement (EI), probability of improvement (PI), and lower confidence bound (LCB) $\text{LCB}_{\overline{\lambda}}(\vv{x}) = -\overline{\lambda}\hat{\mu}(\vv{x}) + \hat{\sigma}(\vv{x})$, where $\hat{\mu}$ and $\hat{\sigma}$ are the mean and standard deviation estimated by surrogate model while $\overline{\lambda}$ is an adjustable ratio between exploitation and exploration. 

In addition, these surrogate models, their hyperparameters, and acquisition functions were chosen because they represent the majority of off-the-shelf options accessible, and are ones that have been widely applied to materials optimization campaigns in the field. Our study provides a comprehensive test across the five datasets in order to reflect how each BO algorithm, resulting from the pairing above, performs across many different materials science design spaces. GP and RF were also selected as examples to specifically illustrate how the differences in implicit distributional assumptions of surrogate model could affect the prediction of the mean and standard deviation when selecting subsequent experiments.


\subsection*{Pool-based active learning benchmarking framework}
\vspace{-5pt}

Within each respective experimental dataset, the set of data points form a discrete representation of ground truth in the materials design space. Figure 2 shows the pool-based active learning benchmarking framework we use to simulate materials optimization campaigns guided by BO algorithms in each materials system. 

The framework has the following properties: 
\begin{enumerate}
  \item It has the traits of an active learning study as it contains a machine learning model that is iteratively refined through subsequent experimental observation selection based on information from previously explored data points. The framework is also adapted for BO, and emphasizes optimization of materials objectives over building an accurate regression model in design space.
  \item It is derived from pool-based active learning. Besides the randomly selected initial experiments, the subsequent experimental observations are selected from the total pool of undiscovered data points ${(\vv{x}, y)} \in D$, whose input features $\vv{x}$ are all made available for evaluation by the acquisition functions. The ground truth in the materials design space was represented with discrete data points over a continuous emulation for the following reasons:
  \begin{enumerate}
       \item Materials design spaces are often not continuous due to experimental resolution limitation and noise in real research scenarios.
       \item Materials datasets do not necessarily capture the domain manifold due to lack of abundant design space sampling density and coverage.
       \item To emulate a domain manifold, selecting of models such as GP introduces smoothness assumptions into the design space, and thus during the benchmarking process could give great advantages to BO algorithms with surrogate models sharing similar assumptions.
  \end{enumerate}
  \item  At each learning cycle of the framework, instead of selecting a larger batch, only one new experiment is obtained. Due to the different dataset sizes across five materials studies, keeping a batch size of one allows us to directly compare the performance of BO algorithms across different materials studies.
\end{enumerate}

\vspace{-5pt}
\begin{figure}[!h]
  \centering
  \includegraphics[width = \columnwidth]{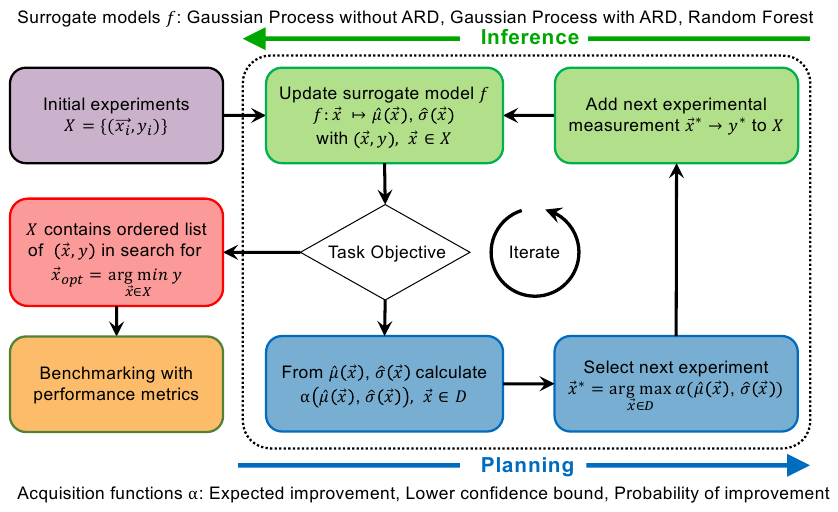}
  \caption{Benchmarking framework including a simulation of BO performing closed-loop optimization with alternating inference and planning stages. \protect$X$ is the iteratively collected sequence of experimental data $(\protect\vv{x}, y)$ during optimization campaign. \protect$D$ is the original pool or total undiscovered set of data from which next experiments are selected. \protect$f$ is the surrogate model used to estimate mean \protect$\hat{\mu}$ and standard deviation \protect$\hat{\sigma}$, which parameterize the acquisition function \protect$\alpha$ to select next experiment $\protect\vv{x}^\text{*}$ to be evaluated.}
  
  \vspace{-13pt}
\end{figure}

Each BO algorithm is evaluated for 50 ensembles with 50 independent random seeds governing the initialization of experiments. The aggregated performances of the BO algorithms derived from 50 averaged runs resulting from 10 random five-fold splits using the 50 original ensembles, is compared against a statistical random search baseline, and we can quantitatively evaluate its performance via active learning metrics defined in the sections below. A detailed description of the framework and the calculation of statistical random baselines can be seen in the Methods section. The simulated materials optimization campaigns were conducted on the Boston University Shared Computing Cluster (SCC). This enabled the parallel execution of multiple campaigns on individual computing nodes. The computing nodes were configured to have access to a maximum of 28 cores (2 fourteen-core 2.4 GHz Intel Xeon E5-2680v4 processors) with a maximum of 256 GB of memory.

\subsection*{Observation of performance through case study on crossed barrel dataset}

While the five datasets covered a breadth of materials domains, the relative performances of tested BO algorithms were observed to be consistent. Figure 3(a) demonsrates the performances of RF, GP ARD (Mat\'ern52), and GP (Mat\'ern52). For the full combinatorial study including all types of GP kernels and acquisition functions, please refer to supplementary information (SI). We showcase the benchmarking results using the crossed barrel dataset \cite{gongora2020bayesian}, which was collected by grid sampling the design space through a robotic experimental system while optimizing the toughness of additive manufactured crossed barrel structure. As for the performance metric, we use
\begin{eqnarray}
\text{Top\%}(i) = \frac{\text{number of top candidates discovered}}{\text{number of total top candidates}} \in [0, 1]
\end{eqnarray}
to show the fraction of the crossed barrel structures with top 5\% toughness that have been discovered by cycle $i = 1, 2, 3,\ldots, N$. $\text{Top\%}$ describes how quickly can a BO guided autonomous experimental system could identify multiple top candidates in a materials design space. Keeping multiple well-performing candidates allows one to not only observe regions in design space that frequently yield high-performing samples but also have backup options for further evaluation should the most optimal candidate fail in subsequent evaluations. There are research objectives related to finding any good materials candidate, yet in those cases, random selection could outperform optimization algorithms due to luck in a simple design space. Our objective of finding multiple or all top tier candidates is more applicable to experimental materials optimization scenarios and suitable for demonstrating the true efficacy and impact of BO. 

\vspace{-2pt}

\begin{figure}
  \centering
  \includegraphics[width = \columnwidth]{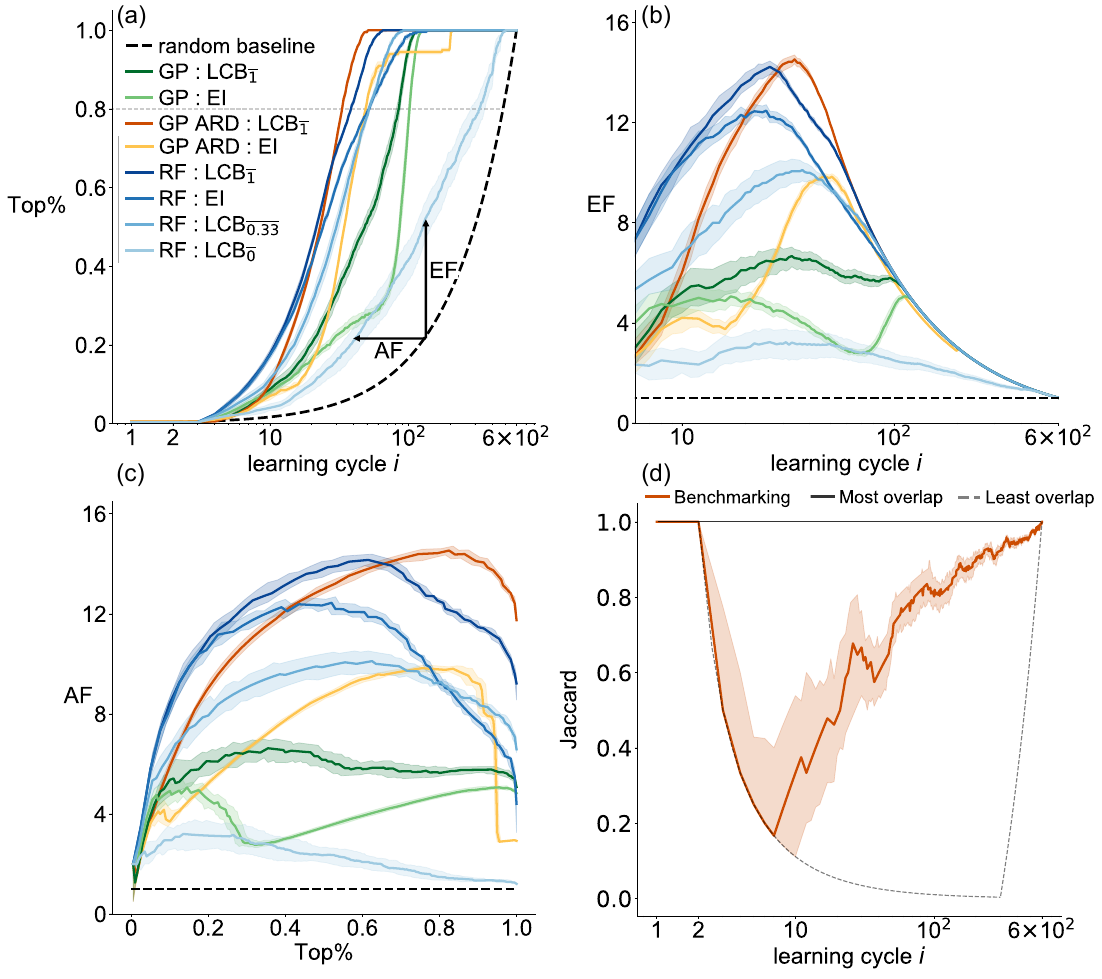}
  \caption{The aggregated performance of BO algorithms on the Crossed barrel dataset measured by a) $\text{Top\%}$ vs. learning cycle $i$ against random baseline, and how b) Enhancement factor $EF$ and c) acceleration factor $AF$ are derived from it. The algorithms with GP as surrogate model are labeled in red, and RF in blue; higher color saturation is correlated with better performance. Variation at each learning cycle is visualized by plotting the median as well as shaded regions representing the 25\textsuperscript{th} to 75\textsuperscript{th} percentile of the aggregated 50-run ensembles. The acquisition functions used are EI and LCB$_{\overline{\lambda}}$ (d) Jaccard similarity index calculated between the optimization campaign sequences of BO algorithms RF: LCB$_{\overline{1}}$ and GP ARD: LCB$_{\overline{1}}$. The median, 25\textsuperscript{th}, 75\textsuperscript{th} percentile of the 50-run ensemble are shown respectively.} 
  \label{fig:perf}
  \vspace{-15pt}
  
\end{figure}

Figure 3(a) illustrates learning rates based on $\text{Top\%}$ metric for RF and GP with multiple acquisition functions. 

\newpage
The following are observed: 
\begin{enumerate}
  \item RF initially excels at lower learning cycles, while GP with ARD takes the lead after $\text{Top\%} = 0.67$. Under the same acquisition function, performance of RF as a surrogate model is on par, if not slightly worse, when compared to the performance of GP with ARD.
  \item Both RF and GP with ARD outclass GP without ARD.
  \item LCB$_{\overline{1}}$, which has equal weights for exploration and exploitation, outperforms other acquisition functions LCB$_{\overline{\lambda}}$ that overly emphasize exploration. LCB$_{\overline{1}}$ also outperformed EI, which is a very popular acquisition function in previous materials optimization studies but has also been known to make excessive greedy decisions \cite{ryzhov2016convergence, hennig2012entropy, frazier2018bayesian}.
 \end{enumerate}
 
Because of this observation, when trying to compare BO algorithms with different surrogate models in this work, the same acquisition function LCB$_{\overline{1}}$ was used. We have also evaluated the performance of BO algorithms using probability of improvement (PI) as acquisition function, but its performance was mostly worse than EI and therefore not the focus of discussion; this observation can be partially attributed to PI only focusing on how likely is an improvement occurs, but not considering how much improvement could be made.

To further quantify the relative performance, we set $\text{Top\%} = 0.8$ as a realistic goal to indicate we have identified 80\% of the structures with top 5\% toughness (Figure 3a). For surrogate models paired with LCB$_{\overline{1}}$, we see that GP with ARD and RF reach that goal by evaluating approximately 30 candidates out of the total of 600, whereas GP without ARD needs about 90 samples. $\text{Top\%}$ rises initially as slowly as the random baseline because the surrogate models suffer from high variance in prediction, having only been trained with a small datasets; $\text{Top\%}$ ramps up very quickly as the model learns to become more accurate in identifying general regions of interest to explore; the rate of learning eventually slows down at high learning cycles because the local exploitation for the global optimum has exhausted most if not all top 5\% toughness candidates, and the algorithms therefore switch to exploring sub-optimal regions. Therefore, it can be assumed that the most valuable regions to examine performance is before each curve reaches $\text{Top\%} = 0.8$ and $\text{Top\%} = 0.8$ can be used as a realistic optimization goal. 
\newline

To quantify the acceleration of discovery from BO, we adapt two other metrics similar to the ones from Rohr et al. \cite{rohr2020benchmarking}. 
Both compared to a statistical random baseline, 

Enhancement Factor(EF)
\begin{eqnarray}
\text{EF}(i) = \dfrac{\text{Top\%}_{\text{BO}}(i)}{\text{Top\%}_{\text{random}}(i)}
\end{eqnarray}

shows how much improvement in a metric one would receive at cycle $i$, 

and Acceleration Factor(AF) \begin{eqnarray}
\text{AF}(\text{Top\%} = a) = \dfrac{i_{\text{BO}}}{i_{\text{random}}}
\end{eqnarray}

is the ratio of cycle numbers showing how much faster one could reach a specific value $\text{Top\%}(i_{\text{BO}}) = \text{Top\%}(i_{\text{random}}) = a \in [0, 1]$. 
The aggregated performance of BO algorithms is further quantified via EF and AF curves in Figure 3(b), 3(c): starting off with small EFs or AFs before the surrogate model gains more accuracy; reaching absolute EF$_\text{max}$ and AF$_\text{max}$ of up to $16\times$. Eventually, the learning algorithms show diminishing returns from an information gain perspective as we progress deeper into our optimization campaigns during pool-based active learning. We observe that for the two BO algorithms both with same acquisition function LCB$_{\overline{1}}$ but different surrogate model GP ARD and RF, they reach EF$_\text{max}$ at different learning cycles and AF$_\text{max}$ at different $\text{Top\%}$, both corresponding to the switch of best performing algorithm around $\text{Top\%} = 0.67$. RF: LCB$_{\overline{1}}$ clearly excels at lower learning cycles, yet GP ARD: LCB$_{\overline{1}}$  takes the lead and would reach $\text{Top\%} = 0.8$ with fewer experiments. Therefore, these results objectively show that optimal BO algorithm selection varies with assigned experiment budget and specific optimization task \cite{rohr2020benchmarking}.


Since we identified two BO algorithms, RF: LCB$_{\overline{1}}$ and GP ARD: LCB$_{\overline{1}}$ , to have similar performance, we wanted to further investigate how similar their optimization paths were in the design space when starting from the same initial experiments. In Figure 3(d), we use Jaccard similarity index to quantify the similarity in optimizations paths. Jaccard similarity, $J = \dfrac{|A\cap B|}{|A\cup B|}$, is the size of the intersection divided by the size of the union of two finite sample sets; specifically in our benchmarking study, using the same 50-ensemble runs that generated Figure 3(a), we can calculate Jaccard similarity value $J(i)$ at each learning cycle $i$, where $A(i)$ is the set of data points sequential collected at each learning cycle during an optimization path guided by BO algorithm GP ARD: LCB$_{\overline{1}}$, and $B(i)$ is that of using RF: LCB$_{\overline{1}}$. As baselines, we have also drawn what the Jaccard similarity value would look like between two optimization paths that begin with the same initial experiments and statistically have least overlap or most overlap. When $i = 1$ or $2$, the same initial experiments are given to the two BO algorithms, and $J = 1$.  When $2 < i < 8$, we can see that the Jaccard similarity value drops as quickly as the statistically least overlapping paths, indicating that despite the fact that GP with ARD and RF were trained on the same initial experiments at the onset, they follow very different paths in the materials design space. This behavior indicates that, despite achieving comparable performance, they exploit the underlying physics differently by virtue of the choice of experiments. 

When $i \geq 8$, $J$ increases with $i$, indicating that the paths chosen by the two algorithms gradually start to have some overlap as they move towards finding crossed barrels structures with high toughness. Recall both algorithms reached $\text{Top\%} = 0.8$ between 30 to 40 learning cycles in Figure 3(a), and between those learning cycles, we observe that $J$ is approximately between 0.6 - 0.65, still considerably far from $J = 1$. This observation shows that while both algorithms have comparable performance in the task of finding crossed barrel structures with good toughness, due to their different choice of surrogate models, their paths towards discovering optimum can differ considerably.

In addition, the Jaccard similarity value does not increase monotonically, and a significant drop can be seen in $J$ such as one around $i = 30$, which coincides with the learning cycles where GP ARD : LCB$_{\overline{1}}$ overtook RF : LCB$_{\overline{1}}$ as best performing algorithm in Figure 3(a). Since the two algorithms used the same acquisition function, this observation shows that while in general the optimization paths of the two algorithms have more overlap overtime, occasional divergence paths still take place because the two algorithms have considerable difference in gathered data used to learn their surrogate models and how their surrogate models predict mean and standard deviation. GP ARD : LCB$_{\overline{1}}$ and RF : LCB$_{\overline{1}}$ started at the same two initial experiments and use the same acquisition function, and the only difference is the surrogate model used. Thus, the divergence and convergence in optimization paths can be again primarily attributed to GP ARD and RF exploiting underlying physics of crossed barrel structure differently. Figure 3(d) highlights the impact of different surrogate model selection beyond final performance, and to provide better guidelines to future research, inspires us to further investigate the role of surrogate models.  

\subsection*{Comparison of performance across datasets}

To further assess the performance of BO, optimization campaigns were conducted for the P3HT/CNT, AgNP, AM ARES, and Perovskite datasets. Across most, if not all the investigated datasets, it was quite observed consistently that the performance of BO algorithms using GP with ARD and RF as surrogate models were comparable, and both outperform those using GP without ARD in most datasets. To illustrate, in Figure 4, we show such relative performance using normalized EF$_\text{max}$ of BO algorithms same acquisition function LCB$_{\overline{1}}$ but with different surrogate models across all five datasets. In each dataset, the BO algorithm with the largest EF$_\text{max}$ had its EF scaled to 1, and the other two BO algorithms showing lower EF$_\text{max}$ were correspondingly scaled, resulting in five sets of column plots. In addition to the observation on relative performance, we also observe that BO algorithms with RF and GP ARD as surrogate model also have plenty overlap between their 25th to 75th percentile across five datasets, further indicating their similarity in performance.   

Notably, EF$_\text{max}$ of the other four datasets datasets were in the $2\times$ to $5\times$ range as seen in Figure S1 from SI, which is much lower than the EF$_\text{max}$ of the crossed barrel dataset in Figure 3(b). The difference in the absolute EF$_\text{max}$ can be attributed to the data collection methodology of the individual datasets. While the crossed barrel dataset was collected using a grid sampling approach, the other four studies were collected along the path of a BO guided materials optimization campaign. Therefore, these four datasets were smaller in size and possessed an intrinsic enhancement and acceleration within their datasets. As a result, it is reasonable that these datasets demonstrate lower EFs, AFs during benchmarking. Nevertheless, the fact that both EFs and AFs are still larger than 1 indicate that further acceleration and enhancement is still possible when given design spaces as parts of optimization paths. Noticeably, the Perovskite dataset had the most intrinsic acceleration because its next experimental choice was guided by BO infused with probabilistic constraints generated from DFT proxy calculations of the environmental stability\cite{sun2021data} of perovskite . As a result, the optimization sequence to be chosen in that study is already narrowed down to a more efficient path from initial experiments to final optimum, making the random baseline to appear arbitrarily much worse. Another interesting observation is how the performance of GP without ARD (isotropic kernels) as surrogate model catches up with GP ARD and RF when the design space has an already "easier" path towards the optimum, further supporting our observations on relative performance. The hypothesis that the lower EF$_\text{max}$ are caused by intrinsic acceleration and enhancement resulting from dataset collection process can be verified by collecting a subset from the uniformed grid sampled crossed barrel dataset. This subset is collected by running BO algorithm GP: EI until all candidates with top 5$\%$ toughness are found, representing an "easier" path towards optimums, and therefore carries intrinsic enhancement and acceleration. We run the same benchmarking framework on this subset,  and observe that while EF$_\text{max}$ is significantly reduced, the relative ranking of the three surrogate models is consistent, as seen again in Figure S1 from SI. Despite the differences described above, all the investigated BO algorithms outperformed the random baseline demonstrating the efficacy of BO in materials optimization campaigns.  

\vspace{-2pt}

\begin{figure}
  \centering
  \includegraphics[width = 10cm]{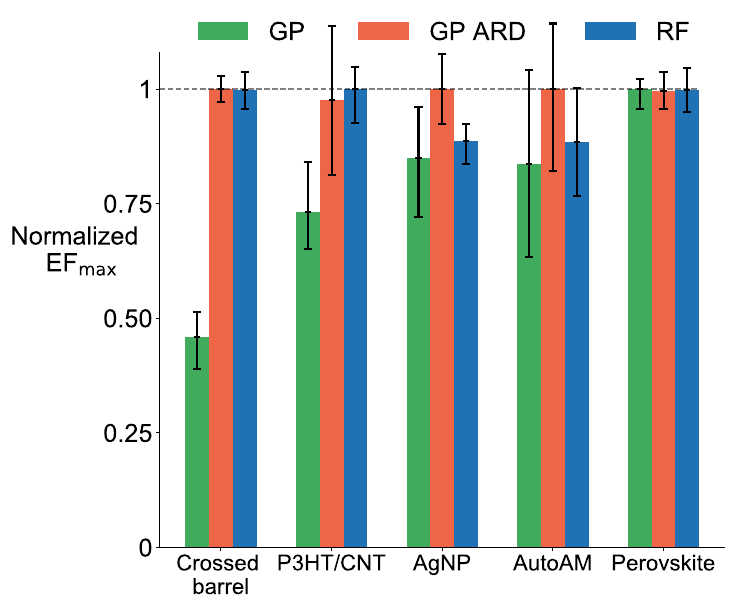}
  \caption{Normalized EF$_\text{max}$ demonstrated by BO algorithms having GP without ARD, GP with ARD, and RF as surrogate models and all using LCB$_{\overline{1}}$ as acquisition function. For each algorithm applied across datasets, the 50\textsuperscript{th} percentile of EF$_\text{max}$ is shown by the barplots, and its 25\textsuperscript{th} and 75\textsuperscript{th} percentile are shown by respective floating bars.}
  
  \vspace{-15pt}
  
\end{figure}

\section*{Discussion}
\vspace{-7.5pt}

In this section, we discuss performance of BO algorithms in the context of autonomous and high-throughput materials optimization. 

\subsection*{Comparison of RF and GP as surrogate models}

While BO algorithms with GP type surrogate models have been extensively used in the field, we observe that the performance of RF as surrogate model is comparable to that of GP. The results heavily suggest that RF is a capable surrogate model to consider besides GP in BO for future HT materials optimization campaigns. 

Before more detailed comparison, an important distinction to make between RF and GP as surrogate models is how they predict mean and standard deviation, which can be attributed to the implicit assumptions when using them as surrogates to represent an unknown ground truth within materials domains. A GP in this work, whether with ARD or not, is essentially a distribution over a materials domain such that any finite selection of data points in this design space results in a multivariate Gaussian density over any point of interest in the space. For the selection of a new data point as next experiment, its predicted mean and standard deviation are all part of a gaussian distribution constructed from previous experiments. Therefore, the predicted means and standard deviations of GPs from their posteriors carry gaussianity assumptions and can be interpreted as statistical predictions based on prior information. Meanwhile, a RF is an ensemble of decision trees, which are trained on the experimental data points collected during optimization and have slight variation due to bootstrapping. For RF, prediction of objective value at a new data point is an aggregated result, most likely the mean \cite{roy2012robustness} as used in this study, of all its decision trees' prediction; similarly, prediction of standard deviation at new data point by RF is the standard deviation of all its decision trees' predictions. Compared to those of GPs, the predicted means and standard deviations of RFs do not have strong distributional assumptions, and can be interpreted as empirical estimates. In short, during prediction, GPs rely on heavy distributional assumptions while RF is distribution free. Such difference between two surrogate models carries over to the values of mean and standard deviation, and together affect the selection of next experiment despite being paired with exact same acquisition function. This important distinction explains the differences of results in Figure 3 and Figure 4. Other factors affecting surrogate model selection are further discussed via the following side-by-side comparisons of GPs and RFs. 

We would first like to discuss the time complexity of these surrogate models. Most commonly in autonomous experimental materials optimization studies, time spent on generating samples is much more significant than that of surrogate model training. However, in our study, there was a noticeable difference in time when benchmarking different BO algorithms, which reveals foreseeable challenges to future research. Across five datasets in this study, starting from the same initial experiments and using the same acquisition function LCB$_{\overline{1}}$, the ratio of average running time to finish benchmarking framework between the three surrogate models is $t_\text{RF}: t_\text{GP}: t_\text{GP ARD} = 1: 1.32: 1.54$. With $n$ as the number of training data, $n_\text{dim}$ as design space dimension, $n_\text{tree}$ as number of decisions trees kept in RF model, in terms of general training time complexity, we have $t_\text{RF} = \mathcal{O}(nlog(n)\cdot n_\text{dim}\cdot n_\text{tree}) < t_\text{GP} = \mathcal{O}(n^\text{3} + n^\text{2}\cdot n_\text{dim})$ \cite{snelson2006sparse, snelson2007flexible, snelson2007local}. The relatively expensive computational complexity of GP model is mostly due to the process of calculating the inverse of an $n$ by $n$ matrix during its training process, and keeping anisotropic kernels certainly adds extra computational time.  For reasonable choice of $n$, $n_\text{dim}$, and $n_\text{tree}$ in materials science research, RF not only can be trained even faster via parallel computing of its decision trees, but also suffer less in performance with increasing $n$ than GP. While the $\mathcal{O}(n^\text{3})$ time complexity of GPs is typically less of a concern when working with smaller datasets, it could quickly become intractable when applied to larger datasets after experimental samples can be generated at unprecedented rate through advancement of automation in materials laboratories. At each learning cycle, time used in synthesis will eventually match with the time used in model retraining and prediction, typically within seconds. Therefore, in the future, if we do not consider the number of experiments as the budget, but instead consider the total experimental time invested, then RF has a potential advantage over GP when we aim to have fast and seamless feedback loop between model and materials experiments. 

We next discuss some properties in the RF, GP, and GP ARD models that would explain the observed differences in performance. The benchmarking effort presented here is unique in its use of abundant experimental materials data with built in noise, often unavoidable in physical science research. The results thus provide a realistic performance evaluation for optimization algorithms in the context of materials research. RF having good performance across the five experimental datasets can be partially attributed to predictions by RF being empirical estimates from its ensemble of decision trees and free of distributional assumptions. The multiple decision trees of RF naturally have low bias and high variance; the aggregation process in RF of different decision trees mitigates this issue, resulting in a model that has relatively low bias and medium variance, and thus are likely more robust to noise and applicable for generalized prediction with unknown assumption \cite{roy2012robustness, breiman2001random}. These properties of RF can be partially observed in Figure 4, where we not only see the performance of RF match that of GP ARD as surrogate model, but also observe the variance of EF$_\text{max}$ for RF is on average lower than those of GP with ARD and GP without ARD. Generally speaking, on one hand, if the ground truth manifold of a design space indeed satisfies the gaussianity assumption of GP, then arguably GP type surrogate models have an advantage in learning a model with low bias and variance. On the other hand, if there were sharp discontinuities, piece-wise constants or changes in orders of magnitude through local regions of materials design space, The decision trees of RF would be able to capture these points accurately and reflect their influences on future predictions via aggregated result. These points are specifically regions of interest to be further investigated by researchers, whether they are new findings or outliers from experiments, but they typically fall out of distributional assumptions of GP and could be smoothed out in the learned model without considerable effort in kernel hyperparameter tuning.

We last discuss the effort required hyperparameter tuning of surrogate model during optimization. Despite HTE drastically increasing the rate of materials data collection, active learning for optimization in new materials domains still requires data collection in a sequential or batched manner. While RF has potentially more hyperparameters such as $n_{\text{tree}}$, max depth, and max split to select, it is less penalized for sub-optimal choice of hyperparameters compared to GP. In this study, across five datasets, as long as sufficient $n_{\text{tree}}$ were used in RF, its performance as surrogate model in BO algorithm has been consistently comparable to that of GP. Other hyperparameters of RF have had less of an impact. Meanwhile, besides the implicit distribution assumption of using a GP type surrogate model, a kernel (covariance function) of GP specifies a specific prior on the domain. Choosing a kernel that is incompatible with the domain manifold could significantly slow down optimization conversion due to loss of generalization, as seen in Figure S2 - S6 from SI, where BO algorithms with GP surrogate models are highly sensitive to kernel selection. For example, Mat\'ern52 kernel analytically requires the fitted GP to be 2 times differentiable in the mean-square sense \cite{rasmussen2010gaussian}, which can be difficult to verify for unknown materials design spaces. Selecting such a kernel could introduce extra domain manifold assumptions to an unfamiliar design space, as we often have limited data to make confident distribution assumptions at optimization onset. Instead of devoting nontrivial experimental budget to optimize the kernels of GP using adaptive kernels \cite{wilson2013gaussian}, automating kernel selection \cite{schlessingerautomated} or keeping a library of kernels available via online learning, RF is an easier off-the-shelf option that allows one to make fewer structural assumptions about unfamiliar materials domains. If a GP surrogate model is still preferred, a Multilayer Perceptron (MLP) kernel \cite{NIPS2009_3628} mimicking neural networks would be suggested as it has comparable performance to other kernels (see Figure S2 - S6 from SI).



%

Admittedly, our benchmarking framework might have given RF a slight advantage by discretizing the materials domain through actively acquiring a new datapoint at each cycle and limiting the choice of next experiments within the pool of undiscovered datapoints. However, the crossed barrel dataset has a sampling density, size, and range within its design space sufficient to cover its manifold complexity. A drawback of RF is that it performs poorly in extrapolation beyond the search space covered by training data, yet in the context of materials optimization campaigns, this disadvantage can be mitigated by clever design of initial experiments, namely using sampling strategies like latin hypercube sampling (LHS). In this way, we can not only preserve the pseudo random nature of selecting initial experiments but also cover a wider range of data in each dimension so that RF surrogate model would not have to often extrapolate to completely unknown regions. Considering our benchmarking results and the comparisons above, RF's relative ease of use paired with the intuitive tuning of LCB's weights to adjust exploration and exploitation forms a BO algorithm suitable for general materials optimization campaigns at early stages. 

\vspace{-5pt}

\subsection*{Benefits of using GP ARD with anisotropic kernels}
\vspace{-7.5pt}



As seen from Figure 4, using anisotropic kernels with GP essentially removes the performance gap between RF and GP with isotropic kernels. with the benefit that the one can exploit the underlying modeling assumptions of the GP ARD for more advanced analysis.




As mentioned earlier, ARD allows us to utilize individual lengthscales for each input dimension $j$ in the kernel function of GP, which are subsequently optimized along learning cycles. These lengthscales in an anisotropic kernel provide a "weight" for measuring relative relevancy of each feature to predicting the objective, i.e. understanding the sensitivity of objective value to each input feature dimension. The reason GP without ARD shows worse performance is as follows: it will have a single lengthscale in an isotropic kernel as scaling parameter controlling GP's kernel function, which is at odds with the fact that each input feature has its distinct contribution to the objective. Depending on how different each feature is in nature, range and units, e.g. solvent composition vs. printing speed, using the same lengthscale in the kernel function for each feature dimension could provide unreliable predictive results. The materials optimization objective naturally has different sensitivities to each input variable, and thus it is rationale then, that the "lengthscale" parameter inside the GP kernel should be independent. In Figure 4, the noticeable improvements of using an anisotropic kernel can be seen in the relative lower performance of GP without ARD compared to that of GP with ARD.  While data normalization can partially alleviate the problem, how it is conducted is highly subject to a researcher's choice, and therefore we would like to raise awareness of the benefits of using GP with anisotropic kernels.


In addition, the lengthscales from the kernels of GP with ARD provides us with more useful information about the input features. These lengthscale values have been used for removing irrelevant inputs \cite{rasmussen2010gaussian}, where high $l_{j}$ values imply low relevancy input feature $j$. In the context of materials optimization, we find the following use of ARD especially useful: ARD could identify a few directions in the input space with specially high “relevance.” This means that if we train GP with ARD on input data with their original units and without normalization, once we extract the lengthscale of each feature $l_{j}$, our GP model in theory should not be able to accurately extrapolate more than $l_{j}$ units away from collected observations in $j^{th}$ dimension. Thus, $l_{j}$ suggests the range of next experiments to be performed in the $j^{th}$ dimension of the materials design space. It also infers a suitable sampling density in each dimension in the experimental setting. When a particular input feature dimension has a relative small $l_{j}$ or large $\dfrac{1}{l_{j}}$, it means that for small change in objective value, we would have a relatively large change in the location within this input feature dimension; thus, the the sampling density or resolution in this dimension should be high enough to capture such sensitivity. In addition, people have considered using information extracted from these lengthscales for even more advanced analysis and variable selection \cite{paananen2019variable}. At the expense of computation time tolerable in the context of materials optimization campaigns, an anisotropic kernel provides not only a better generalizable GP model but also useful information in analyzing input feature relevancy at each learning cycle. For the above mentioned reasons, it would be great practice for researchers to emphasize their use of GP with anisotropic kernels as surrogate models during materials optimization campaigns. 

In conclusion, we benchmarked the performance of BO algorithms across five different experimental materials science domains. We utilize a pool-based active learning framework to approximate experimental materials optimization processes, and adapted active learning metrics to quantitatively evaluate the enhancement and acceleration of BO for common research objectives. We demonstrate that when paired with the same acquisition functions, RF as surrogate model can compete with GP with ARD, and both outperform GP without ARD. In the context of autonomous and high-throughput experimental materials research, we discuss the differences in implicit distributional assumptions of surrogate models and the benefits of using GP with anisotropic kernels. We provide practical insights on surrogate model selection for materials optimization campaigns, and also offer open source implementation of benchmarking code and datasets to support future algorithmic development. 

Establishing benchmarks for active learning algorithms like BO across a broad scope of materials systems is only a starting point. Our observations demonstrate how the choice of active learning algorithms has to adapt to their applications in materials science, motivating more efficient ML guided closed-loop experimentation, and will likely directly result in a larger number of successful optimization of materials with record breaking properties. The impact of this work can be extended to not only other materials systems, but also a broader scope of scientific studies utilizing closed-loop and high-throughput research platforms. Through our benchmarking effort, we hope to share our insights with the field of accelerated materials discovery and motivate a closer collaboration between ML and physical science communities.

\newpage
\section*{Methods}
\vspace{-7.5pt}
\subsection*{Prediction by surrogate models and acquisition functions}
\vspace{-7.5pt}

In order to estimate the mean $\hat{\mu}(\vv{x}_\text{*})$ and standard deviation $\hat{\sigma}(\vv{x}_\text{*})$ of predicted objective value at a previously undiscovered observation $\vv{x}_\text{*}$ in design space:

For a Gaussian process (GP), it assumes a prior over the design space that is constructed from already collected observations $(\vv{x_{i}}, y_{i})$, $i = 1,2,...,n$. This prior is the source of implicit distributional assumptions, and when an undiscovered new observation $(\vv{x}_{\text{*}}, y_{\text{*}})$ is being considered during noisy setting ($\sigma = 0.01$), the joint distribution between the objective values of collected data  $\vv{y} \in \mathcal{R}^n$ and $y_{\text{*}}$ is 
\begin{eqnarray}
\begin{bmatrix} \vv{y} \\ y_{\text{*}} \end{bmatrix} \sim \mathcal{N} \begin{pmatrix} 0, && \begin{bmatrix} K + \sigma^2I && K_{\text{*}}^{T} \\ K_{\text{*}} && K_{\text{**}} \end{bmatrix} \end{pmatrix}. 
\end{eqnarray}

K is the covariance matrix of the input features $X = \{\vv{x_{i}} | i = 1,2,...,n \}$; $K_{\text{*}}$ is the covariance between the collected data and new input feature $\vv{x}_{\text{*}}$; $K_{\text{**}}$ is the covariance between the new data. For each of the covariance matrices, $K_{pq} = k(\vv{x}_{p}, \vv{x}_{q})$, where $k$ is the kernel function, whether isotropic or anisotropic, used in GP. Then from the posterior, we have estimates 
\begin{eqnarray}
\hat{\mu}(\vv{x}) = y_{\text{*}} = K_{\text{*}}[K +\sigma^2I]^{-1}\vv{y}
\end{eqnarray}
and covariance matrix
\begin{eqnarray}
cov(y_{\text{*}}) = K_{\text{**}} - K_{\text{*}}[K +\sigma^2I]^{-1}K_{\text{*}}^{T}
\end{eqnarray}

The standard deviation value $\hat{\sigma}(\vv{x})$ can be obtained from the diagonal elements of this covariance matrix.

For a random forest (RF),  let $\hat{h}_{\text{k}}(\vv{x}_\text{*})$ denote the prediction of objective value from the $k^\text{th}$ decision tree in the forest, $k = 1,2,...,$ $n_\text{tree}$, 
then 
\begin{eqnarray} 
\hat{\mu}(\vv{x}_\text{*}) = \dfrac{1}{n_\text{tree}}\sum_{k=1}^{n_\text{tree}}\hat{h}_{\text{k}}(\vv{x}_\text{*}) \end{eqnarray} 
and 
\begin{eqnarray} \hat{\sigma}(\vv{x}_\text{*}) = \sqrt{\dfrac{\sum_{k=1}^{n_\text{tree}}(\hat{h}_{\text{k}}(\vv{x}_\text{*}) - \hat{\mu}(\vv{x}_\text{*}))^2}{n_\text{tree}}}
\end{eqnarray}
The median or other variations could also be used in future studies to aggregate the predictions for potential improvement in robustness \cite{roy2012robustness}. 


We tested three acquisition functions in our study, including expected improvement (EI), probability of improvement (PI), and lower confidence bound (LCB).
\begin{eqnarray}
\text{EI}(\vv{x}) = (y_{\text{best}} - \hat{\mu}(\vv{x}) - \xi)\cdot\Phi(Z) + \hat{\sigma}(\vv{x})\varphi(Z)
\end{eqnarray}
\begin{eqnarray}
\text{PI}(\vv{x}) = \Phi(Z)
\end{eqnarray}
where 
\begin{eqnarray}
Z = \dfrac{y_{\text{best}} - \hat{\mu}(\vv{x}) - \xi}{\hat{\sigma}(\vv{x})}
\end{eqnarray}
$\hat{\mu}$ and $\hat{\sigma}$ are estimated mean and standard deviation by surrogate model; $y_{\text{best}}$ is best discovered objective value within all collected values so far; $\xi = 0.01$ is jitter value that can slightly control exploration and exploitation; $\Phi$ and $\varphi$ are the cumulative density function and probability density function of a normal distribution.
\begin{eqnarray}
\text{LCB}_{\overline{\lambda}}(\vv{x}) = -\overline{\lambda}\hat{\mu}(\vv{x}) + \hat{\sigma}(\vv{x})
\end{eqnarray}
where $\overline{\lambda}$ is a adjustable ratio between exploitation and exploration. 

\subsection*{Pool-based active learning framework}
\vspace{-7.5pt}

As seen in Figure 2, to approximate early stage exploration during each optimization campaign, $n = 2$ initial experiments are drawn randomly with no replacement from original pool $D = \{(\vv{x_{i}}, y_{i})|i=1,2,\ldots,N\}$ and add to collection $X = \{(\vv{x_{i}}, y_{i})|i=1,2,\ldots,n\}$. During planning stage, surrogate model $f$ is used to estimate the mean $\hat{\mu}(\vv{x})$ and standard deviation $\hat{\sigma}(\vv{x})$. We then evaluate the acquisition function values $\alpha(\hat{\mu}(\vv{x}), \hat{\sigma}(\vv{x}))$ for each remaining experimental action $\vv{x}\in D$ in parallel. At each cycle, action $\vv{x}^*= \argmaxA_x \alpha(\vv{x})$ will be selected as next experiment. During inference stage, after selecting action $\vv{x}^*$, the corresponding sample observation $y^*$ is obtained, and $(\vv{x}^*, y^*)$ is added to $X$ and removed from set $D$. The new observation $(\vv{x}^*, y^*)$ is incorporated into the surrogate model. The sequential alternation between planning and inference is repeated until undiscovered data points run out.

\subsection*{Statistical baselines}

In Figure 3 amd 4, we have introduced some statistical baselines when benchmarking the performance of BO algorithms with a pool-based active learning framework.

For the random baseline in Figure 3(a), assuming a total pool of $N$ data points and the number of good materials candidates $M = 0.05N$, at cycle $i = 1$, expected probability of finding a good candidate is $P(1) = 0.05$ and expected value of $\text{Top}\%(1) = \dfrac{1 \cdot P(1)}{M} = 0.0016$. 

Then at cycle $i = 2, 3, ..., N$,  there is 

\begin{eqnarray}
P(i) = \dfrac{M - \sum_{n = 1}^{i-1} P(n)}{N -i}
\end{eqnarray} 
and 
\begin{eqnarray} \text{Top}\%(i) = \dfrac{\sum_{n = 1}^{i} P(n)}{M}
\end{eqnarray}

In Figure 3(d), between two optimization paths starting with the same two initial data points:

\begin{enumerate}
    \item The statistically most overlap happens when two paths are identical, resulting in $J(i) = 1$, $i = 1, 2,..., N$;
    
    \item The statistically least overlap happens when the two follow drastically different paths until they run out of data points undiscovered by both algorithms, resulting in 
    \begin{eqnarray}
        J(i) = \begin{cases} 
                1 & 1 \leq x\leq 2 \\
                \dfrac{1}{i - 1} & 3 \leq i\leq \dfrac{N}{2}+1 \\
                \dfrac{2i - N}{N} & \dfrac{N}{2}+2 \leq i\leq N
                \end{cases}
    \end{eqnarray}
\end{enumerate}

\vspace{-10pt}

\section*{Data availability}
The five experimental datasets in the current study is available are the following GitHub repository \cite{benchmarkingGithub}: https://github.com/PV-Lab/Benchmarking.

\section*{Code availability}
The code for pool-based active learning framework and visualization in the current study are available in the following GitHub repository \cite{benchmarkingGithub}: https://github.com/PV-Lab/Benchmarking.

\bibliographystyle{unsrt}

\end{document}